\documentstyle[12pt,emulateapj]{article}

\slugcomment{To appear in \apj}

\lefthead{Nitta et al.}
\righthead{Self-similar Evolution of Fast Reconnection}

\begin{document}

\title{Self-similar solution of fast magnetic reconnection: Semi-analytic study of inflow region}

\author{S. Nitta \altaffilmark{1}}
\affil{Department of Astronomical Science, The Graduate University for Advanced Studies, Osawa 2-21-1, Mitaka 181-8588, Japan}
\authoremail{snitta@th.nao.ac.jp}

\author{S. Tanuma \altaffilmark{2}}
\affil{Solar-Terrestrial Environment Laboratory, Nagoya University, 3-13 Honohara, Toyokawa, Aichi 442-8507, Japan}
\authoremail{tanuma@stelab.nagoya-u.ac.jp}

\and

\author{K. Maezawa}
\affil{Institute of Space and Astronautical Science 3-1-1 Yoshinodai, Sagamihara 229-8510 Japan}
\authoremail{maezawa@stp.isas.ac.jp}

\altaffiltext{1}{Division of Theoretical Astrophysics, National Astronomical Observatory of Japan, Osawa 2-21-1, Mitaka 181-8588, Japan}

\altaffiltext{2}{Current address: Kwasan and Hida Observatories, Kyoto University, Yamashina, Kyoto 607-8417 Japan}

\begin{abstract}
\end{abstract}

An evolutionary process of the fast magnetic reconnection in ``free space'' which is free from any influence of outer circumstance has been studied semi-analytically, and a self-similarly expanding solution has been obtained. The semi-analytic solution is consistent with the results of our numerical simulations performed in our previous paper (see Nitta et al. 2001). This semi-analytic study confirms the existence of self-similar growth. On the other hand, the numerical study by time dependent computer simulation clarifies the stability of the self-similar growth with respect to any MHD mode. These results confirm the stable self-similar evolution of the fast magnetic reconnection system. 

\keywords{ Earth---MHD---Sun: flares---ISM: magnetic fields}

\section{Introduction}
\label{sec:intro}

Magnetic reconnection is the most probable candidate of the energy conversion process for solar flares. Such an explosive event in plasma systems is very common in astrophysical problems, e.g., geomagnetospheric substorms, YSO flares and origins of galactic ridge X-ray emission. However, entire macroscopic structure of evolutionary reconnection is still unclear. 

Actual magnetic reconnection in astrophysical systems usually grows in a huge dynamic range in its spatial dimension. The initial scale of the reconnection system can be defined by the initial current sheet thickness. Let us assume it to be of the order of the ion Larmor radius ($\sim 10^0$[m] for solar corona). Eventually, the reconnection system will develop into the scale comparable to the initial curvature radius of the magnetic field lines ($\sim 10^7$[m] for typical magnetic loop of solar corona). The dynamic range of the spatial scale in evolution is therefore extremely large ($\sim 10^7$ for solar flares). For the geomagnetospheric substorms, their dynamic range of growth is also large ($\sim 10^4$ for substorms). Such a very wide dynamic range of growth suggests that the evolution of the magnetic reconnection should be treated as a development in free space, and individual conditions of the outer circumstances of evolution do not affect the evolution of magnetic reconnection, at least in the initial stage. 

The numerical study of magnetic reconnection in free space has been performed by our group (see Nitta, Tanuma, Shibata and Maezawa 2001), and we have discovered a self-similar evolutionary process of the fast reconnection. In our work, a stable self-similar growth of the reconnection system was shown. We here note the definition of the self-similar growth as following: The solution at arbitrary time is the same with the solution at any other time if we change the scale of physical quantities (see section \ref{sec:zoom-out}). The spatial scale is expanding in proportion to the time. Such self-similar evolution is expected to continue unlimitedly in our case. However, simulated dynamic range of self-similar growth was restricted ($\sim 10^3$) owing to the restriction of computer memory and CPU time. 

The reconnection model with a pair of slow-mode shocks was first presented by Petschek (1964). In this model, the magnetic energy is mainly converted at the slow-mode shocks. Therefore, the time scale of energy conversion is determined by that of MHD wave propagation, and, as such, is much shorter than that of the simple diffusion model (Tajima \& Shibata 1997) and the Sweet-Parker model (Sweet 1958, Parker 1963). According to the Petschek model, the reconnection rate is almost independent of the magnetic Reynolds number ($\propto \log R_m \sim {R_m}^0$). Hence, we call this quick magnetic energy conversion ``fast reconnection''. 

We should note, however, that the question of what controls the reconnection rate is still open. Concerning this point, there have been two different general ways of producing reconnection. One assumes that external boundary conditions should control the reconnection, so that the resistivity has no effective influence on the energy conversion (see, e.g., Petschek 1964, Priest \& Forbes 1986 for theoretical studies, and Sato \& Hayashi 1979 for numerical simulations). Another way has been proposed in a series of numerical works originated by Ugai \& Tsuda (Ugai \& Tsuda 1977, 1979, Tsuda \& Ugai 1977, and recently, Ugai 1999). In their simulation, they put a localized finite resistivity in the current sheet to represent an anomalous resistivity that may exist in plasmas. This resistivity acts as a trigger to start the magnetic reconnection. Once this resistivity is put in, the reconnection starts and evolves self-consistently, and forms a fast-reconnection system with the Petschek-type structure (a pair of slow-mode shocks is formed along the current sheet). 

However, even in the previous numerical studies aimed at clarification of the time evolution of fast reconnection, the evolution could not be followed for a long time. This is mainly owing to the finite size of the simulation box, hence the application of the results has been limited to spatial scales typically, say, hundred times the spatial scale of the diffusion region. We are interested in evolutionary process in a free space without any influence of the outer circumstance. In such a system, the evolution and resultant structure would be quite different from these previous models. The key process of the self-similar reconnection is summarized as following. 

We suppose a two-dimensional equilibrium state with anti-parallel magnetic field distribution, as in the Harris solution. When magnetic diffusion takes place in the current sheet by some localized resistivity, magnetic reconnection will occur, and a pair of reconnection jet is ejected along the current sheet. This causes a decrease of the total pressure near the reconnection point. Such information propagates outward as the rarefaction wave. In a low-$\beta$ plasma ($\beta \ll 1$ in the region very distant from the current sheet [asymptotic region]; as typically encountered in astrophysical problems), the propagation speed of the fast-magnetosonic wave is isotropic, and is much larger than that of other wave modes. Thus, the information about the decreasing total-pressure propagates almost isotropically as a fast-mode rarefaction wave (hereafter we call it FRW) with a speed almost equal to the Alfv\'{e}n speed $V_{A0}$ in the asymptotic region. Hence, the wave front of FRW (hereafter we call it FRWF) has a cylindrical shape except near the point where FRWF intersect the current sheet. When FRWF sufficiently expands, the initial thickness of the current sheet becomes negligible comparing with the system size $V_{A0} t$, where $t$ is the time from the onset of reconnection. In such a case, there is only one characteristic scale, i.e., the radius of FRWF ($V_{A0} t$), which increases linearly with proceeding of the time. This is just the condition for the self-similar growth. 

In this paper, our attention is focused on the analytic study of self-similar stage of magnetic reconnection in free space. An analytic approach has the following importance. 

1) By finding the analytic solution for the self-similar growth, we can verify the possibility of the self-similar evolution more rigorously. We set the initial thickness of the current sheet $D$. The self-similar stage is realized in the limit of $V_{A0} t \gg D$. However, in the computer simulation, we can only perform calculation for a finite duration owing to technical reasons. Hence, we will never reach the exact self-similar stage. But, if we can verify that the result of the numerical simulation is very similar to the analytic self-similar solution, we can accept that the result presented by our numerical simulation is truly a self-similar evolution. 

2) Existence of the analytic solution ensures that the self-similar growth will continue unlimitedly. This is important because, in the computer simulation, we cannot continue the calculation more than the case we performed in the previous paper owing to technical reasons. Of course, we should note that when the system sufficiently grows to a scale similar to the initially imposed system size (e.g., the scale of the flux tube for the case of solar flares), this self-similar evolution must be modified by its circumstance. We should note that such analytic treatment cannot ensure the stability of the self-similar evolution, while it is ensured by our numerical simulation. Thus the analytic treatment and numerical simulation are complementary to each other. 

This paper is organized as follows. In section 2, we introduce a special coordinate system called ``the zoom-out coordinate''. In this coordinate, the real spatial scale is shrinking linearly as the time proceeds. If we choose the origin of time appropriately, self-similar expansion can be a stationary solution in this coordinate. We should modify the MHD equations in a relevant form in the zoom-out coordinate. In section 3, we review basic assumptions which will be made in this work. We study magnetic reconnection in the limit of very small reconnection rate, and adopt the method of perturbative expansion by using the reconnection rate as a small perturbation parameter. The initial equilibrium is treated as the zeroth order (unperturbed) situation. We solve the linearized problem for the first order quantities, which show the variation from the initial state. We can rearrange the first order equations to a single 2nd order partial differential equation for magnetic vector potential. This procedure is called the Grad-Shafranov approach. We will see that this equation is elliptic in our problem. Using the SOR routine for solving the Grad-Shafranov equation, we obtain a self-similar solution. The boundary condition used in this work is discussed at the end of section 3. In section 4, properties of this self-similar solution are discussed. In section 5, we summarize our results, and discuss the properties of this self-similar evolution model.

\section{MHD equations in Zoom-out coordinate}

\subsection{Basic assumptions}
We focus our attention on the inflow region of the reconnection system which is spontaneously evolving without any influence of the outer boundary conditions. Macroscopic aspects of the evolution are well described by fluid approximation, hence, we can use MHD equations for this research. In the fast reconnection models, the resistivity plays an important role only at the diffusion region which is in the current sheet, thus, we can neglect the resistive effects in the inflow region. In usual astrophysical plasma systems, the plasma-$\beta$ value is very small (typically $\beta \sim 10^{-2}$ for solar corona). Hence, we can assume the system is filled with non-resistive, and pressure-free (cold) plasmas (of course, except inside of the current sheet). 

We study the reconnection starting from an equilibrium state of two-dimensional current sheet system with anti-parallel magnetic distribution, as like in the Harris solution. As discussed in the previous section, the self-similar stage is realized in a very late period ($V_{A0} t \gg D$). This means that the initial current sheet thickness $D$ becomes negligible in comparison with the system size $V_{A0} t$. We can estimate the only one system size by the fast-mode wave transit scale $V_{A0} t$. This is equivalent to assuming an infinitesimally thin current sheet. 

In the case of the Petschek-type reconnection, the resultant reconnection rate is the order of $10^{-2}$ which is consistent with our numerical simulation for self-similar fast reconnection (Nitta et al. 2001). We may expect a rather small reconnection rate in many cases. This means that variation of physical quantities due to reconnection from initial equilibrium is very small, and we can treat such variation by a perturbation method.

\subsection{Zoom-out coordinates}
\label{sec:zoom-out}
``Self-similar solution'' may be regarded as a stationary solution in the ``zoom-out coordinate'' which is defined as 

\begin{equation}
\mbox{\boldmath $r$}' \equiv \mbox{\boldmath $r$}/(V_{A0} t) \ ,\label{eq:n-r}
\end{equation}

where {\boldmath $r$} is the position vector from the reconnection point in conventional coordinate system (here after we call this the ``fixed coordinate'') and {\boldmath $r$}$'$ is the position vector in the zoom-out coordinate system. One must note that $V_{A0}$ in equation (\ref{eq:n-r}) represents the Alfv\'{e}n speed at the asymptotic region far from the current sheet, and is constant throughout the evolution. $\mbox{\boldmath $r$}'$ can be expressed as $\mbox{\boldmath $r$}'=x \mbox{\boldmath $i$}+y \mbox{\boldmath $j$}+z \mbox{\boldmath $k$}$, where $x$-axis is parallel to both the current sheet and initial anti-parallel magnetic field, $y$-axis is perpendicular to the current sheet, and $z$-axis is parallel to the current sheet, but is perpendicular to the initial magnetic field (hence, $\partial_z$=0 in this two-dimensional problem). The origin of this coordinate system is at the reconnection point which is chosen to be the center of the current sheet (see, section 3.1 of Nitta et al. 2001). 

In order to avoid ambiguity of the term ``self-similar evolution'', let us consider the following situation. Let the distribution of a physical quantity, say $Q$, to be time varying in the fixed coordinate (i.e., $Q=f[\mbox{\boldmath$r$}, t]$). We measure the distribution of $Q$ in the zoom-out coordinate. If $Q=g(\mbox{\boldmath$r$}')$ is independent of time, we can say that the distribution of $Q$ is self-similarly expanding at a constant speed in the fixed coordinate. In our case, $Q$ denotes any of the following variables: mass density $\rho$, gas pressure $P$, flow velocity $\mbox{\boldmath$v$}$, magnetic field $\mbox{\boldmath$B$}$, and any combination of these quantities (e.g., magnetic pressure, total pressure, etc.). 

Let us express the MHD variables in a non-dimensional form. Note that we give a general formalism here, hence gas pressure is remained in the following equations (when we solve these equations actually, we will take the cold limit and neglect the pressure terms). We adopt the following normalization in this paper, 
\begin{eqnarray}
\mbox{\boldmath $v$}&=&V_{A0} (\mbox{\boldmath $v$}'+\mbox{\boldmath $r$}') \label{eq:n-v}\\
\rho&=&\rho_0 \cdot \rho'(\mbox{\boldmath $r$}')\\
\mbox{\boldmath $B$}&=&V_{A0} \sqrt{\mu_0 \rho_0} \cdot \mbox{\boldmath $B$}'(\mbox{\boldmath $r$}')\\
P&=&\beta/2 \cdot \rho_0 {V_{A0}}^2 \cdot P'(\mbox{\boldmath $r$}') \label{eq:n-P}
\end{eqnarray}
where $\mu_0$ is the magnetic permeability in vacuum, $\beta \equiv (C_{s0}/V_{A0})^2$ with $C_{s0}$ being the sound speed at the asymptotic region. Note that we use SI units throughout this paper.

\subsection{Non-dimensional MHD equations in zoom-out coordinate}
By using above normalization, the MHD equations in the zoom-out coordinate can be put into the following non-dimensional form (see appendix A), 
\begin{eqnarray}
&\mbox{\boldmath $\nabla$}' \cdot (\rho' \mbox{\boldmath $v$}')=-2\rho' \label{eq:cont-z}\\
&\mbox{\boldmath $\nabla$}' \cdot [\rho' \mbox{\boldmath $v$}' \mbox{\boldmath $v$}'-\mbox{\boldmath $B$}' \mbox{\boldmath $B$}'+\mbox{\boldmath $I$}(\beta P'+B'^2)/2]=-3 \rho' \mbox{\boldmath $v$}' \label{eq:mom-z}\\
&\mbox{\boldmath $\nabla$}' \cdot (\mbox{\boldmath $v$}' \mbox{\boldmath $B$}'-\mbox{\boldmath $B$}' \mbox{\boldmath $v$}')=-\mbox{\boldmath $B$}' \label{eq:ind-z}\\
&P' \rho'^{-\gamma}=1 \label{eq:pol-z}
\end{eqnarray}
where {\boldmath $I$} is the unit tensor and $\gamma$ is the specific heat ratio. 

The left-hand side of each equation is very similar to its counter part in ordinary MHD equations, but strange source terms appear in the right-hand side. These source terms appear as apparent effects in the zoom-out coordinate. We should note that our main focus is on the structure of the inflow region, which is located outside of the diffusion region, and upstream of the slow shock as in the Petschek model. The resistivity does not play any important role in the inflow region. Thus, we can use ideal (non-resistive) MHD equations. For simplicity, we adopt the polytropic relation instead of the full energy equation. The flow does not experience a violent entropy production in the inflow region (up-stream side of the slow shock). Thus, the polytropic variation is a good approximation in the inflow region.

\section{Equation for self-similar solution}

\subsection{Perturbative expansion}

We are treating here a slow energy conversion induced by magnetic reconnection. The word ``slow'' used here means very small reconnection rate, and should not be confused with that in the term ``slow reconnection'', which usually means an explicit dependence of the reconnection rate on the magnetic Reynolds number. As we found numerically, the self-similar reconnection we wish to study is ``fast'' reconnection in the sense that the reconnection rate does not depend upon the magnetic Reynolds number (at most, it depends logarithmically). 

In such a case, deviation owing to reconnection from the initial state is very small in the inflow region. Hence we can adopt the perturbative expansion method. We treat the quantities of the initial equilibrium state to be the zeroth order, and any variation from it should be treated as the first order quantity. In this paper, our attention is focused to fundamental properties of the self-similar reconnection; thus we treat up to the first order variation. We can recognize that the small expansion parameter in this work is the reconnection rate. 

We assume an anti-parallel two dimensional magnetic distribution with infinitesimally thin current sheet. Note that the self-similar stage is established in the case that there is no fixed proper length in the system. However, the system does have a fixed scale: the initial thickness $D$ of the current sheet is finite. Thus, such self-similar stage is realized only if $V_{A0} t \gg D$ (see section 2 of Nitta et al. 2001). 

This initial equilibrium state (uniform distribution) is expressed as 

\begin{eqnarray}
\mbox{\boldmath $B$}'_0&=&\mbox{\boldmath $i$} \label{eq:b0}\\
\mbox{\boldmath $v$}'_0&=&-\mbox{\boldmath $r$}' \label{eq:v0}\\
P_0'&=&1 \label{eq:p0}\\
\rho_0'&=&1 \label{eq:rho0}
\end{eqnarray}

in the upper half plane $y>0$. The quantities having suffix $0$ are the zeroth order quantities. 

The first order quantities represent the deviation from the initial equilibrium. By substituting the expansion form of each quantity 
\begin{eqnarray}
\mbox{\boldmath $B$}'&=&\mbox{\boldmath $B$}'_0+\mbox{\boldmath $B$}'_1 \label{eq:B0+1}\\
\mbox{\boldmath $v$}'&=&\mbox{\boldmath $v$}'_0+\mbox{\boldmath $v$}'_1 \label{eq:v0+1}\\
P'&=&P'_0+P'_1 \label{eq:P0+1}\\
\rho'&=&\rho'_0+\rho'_1 \label{eq:rho0+1}
\end{eqnarray}
together with equations (\ref{eq:b0}) - (\ref{eq:rho0}) into the MHD equations (\ref{eq:cont-z})-(\ref{eq:pol-z}) in the zoom-out coordinate, we obtain the equations for each order of magnitude. The zeroth order equations are satisfied automatically since the zeroth order quantities show the uniform equilibrium state (see appendix B). 

In order to solve the reconnection in a low $\beta$ plasma, let us take here the cold limit $\beta \rightarrow 0$. 

We obtain the first order equations (see appendix B), 
\begin{eqnarray}
&\mbox{\boldmath $\nabla$}' \cdot \mbox{\boldmath $v$}_1'-\mbox{\boldmath $r$}' \cdot \mbox{\boldmath $\nabla$}' \rho_1'=0 \label{eq:lin-con}\\
&-\mbox{\boldmath $r$}' \cdot \mbox{\boldmath $\nabla$}' \mbox{\boldmath $v$}_1'-\mbox{\boldmath $i$} \cdot \mbox{\boldmath $\nabla$}' \mbox{\boldmath $B$}_1'+\mbox{\boldmath $\nabla$}' (\mbox{\boldmath $i$} \cdot \mbox{\boldmath $B$}_1')=0 \label{eq:lin-mom}\\
&\mbox{\boldmath $r$}' \cdot \mbox{\boldmath $\nabla$}' \mbox{\boldmath $B$}_1'+\mbox{\boldmath $i$} \cdot \mbox{\boldmath $\nabla$}' \mbox{\boldmath $v$}_1'-\mbox{\boldmath $r$}' \cdot (\mbox{\boldmath $\nabla$}' \rho_1') \mbox{\boldmath $i$}=0 \label{eq:lin-ind}\\
&P_1'=\gamma \rho_1' \ . \label{eq:lin-pol}
\end{eqnarray}
Note that even in the cold limit, the dimension-less gas pressure $P'$ has a finite value (see equation [\ref{eq:n-P}]). When we obtain the value of dimension-less density $\rho'$, we can easily obtain $P'$ by equations (\ref{eq:P0+1}) and (\ref{eq:lin-pol}). By solving these first order equations, we can obtain the self-similar solution.

\subsection{Linearized Grad-Shafranov equation}
We adopt the method of the Grad-Shafranov (G-S) equation. Above basic equations (\ref{eq:lin-con})-(\ref{eq:lin-pol}) show that every physical quantity can be expressed as a function of the $z$-component $A_1'$ of magnetic vector potential. Basic physical quantities are related to $A_1'$ in the following way (see appendix C). 
\begin{eqnarray}
\mbox{\boldmath $B$}_1'&=&\mbox{\boldmath $\nabla$}' \times A_1'\mbox{\boldmath $k$} \label{eq:B1}\\
v_{1x}'&=&0 \label{eq:v1x}\\
v_{1y}'&=&x \frac{\partial A_1'}{\partial x}+y \frac{\partial A_1'}{\partial y}-A_1' \label{eq:v1y}\\
\rho_1'&=&\frac{\partial A_1'}{\partial y} \label{eq:rho1}\\
P_1'&=&\gamma \frac{\partial A_1'}{\partial y} \label{eq:P1}
\end{eqnarray}
where the suffix $x$ or $y$ denotes the $x$- and $y$-component of vector quantities, respectively. 

By substituting these functional form to the above linearized MHD equation (\ref{eq:lin-con})-(\ref{eq:lin-pol}), we can obtain the G-S equation for this linearized situation (see appendix C), 
\begin{equation}
(1-x^2) \frac{\partial^2 A_1'}{\partial x^2}-2xy\frac{\partial^2 A_1'}{\partial x \partial y}+(1-y^2) \frac{\partial^2 A_1'}{\partial y^2}=0 \ . \label{eq:GS}
\end{equation}

In our zoom-out coordinate, the Alfv\'{e}n wavefront (or the fast-mode wavefront in the cold plasma limit $\beta \rightarrow 0$; FRWF)  emitted from the reconnection point is a unit circle. Hence, we solve the G-S equation inside the FRWF. 

We should note that equation (\ref{eq:GS}) is a Tricomi type second order partial differential equation. We call the region $|\mbox{\boldmath $r$}'|<1$ of the upper half plane ($y>0$) as the region $R$, and $|\mbox{\boldmath $r$}'|>1$ as $\bar{R}$. The equation is elliptic in $R$, and hyperbolic in $\bar{R}$. The region $R$ is affected by the FRW emitted from the diffusion region. In $\bar{R}$, there is no difference from the initial equilibrium state (uniform distribution), because no signal propagates here yet. We do not need to solve the region $\bar{R}$. Hence, in the following, we solve the elliptic equation (\ref{eq:GS}) for $R$ under boundary conditions on FRWF and on the bottom boundary ($y=0$, which is the junction surface to the reconnection jet). 

It is convenient to rewrite the G-S equation (\ref{eq:GS}) and other relations (\ref{eq:v1x})-(\ref{eq:P1}) in the polar coordinate ($r$, $\theta$), because the boundary of FRWF has a circular shape. We denote by $r \equiv \sqrt{x^2+y^2}$ the distance from the reconnection point $x=y=0$ and by $\theta$ the angle in the $x-y$ plane from $x$-axis (i.e., $\tan \theta = y/x$). The G-S equation and other relations are rewritten as 

\begin{equation}
r^2 (1-r^2) \frac{\partial^2 A_1'}{\partial r^2}+\frac{\partial^2 A_1'}{\partial \theta^2}+r\frac{\partial A_1'}{\partial r}=0 \ , \label{eq:GS-pol}
\end{equation}

and 

\begin{eqnarray}
v_{1x}'&=&0 \label{eq:v1x-pol}\\
v_{1y}'&=&r \frac{\partial A_1'}{\partial r}-A_1' \label{eq:v1y-pol}\\
\rho_1'&=&\sin \theta \frac{\partial A_1'}{\partial r}+\frac{\cos \theta}{r} \frac{\partial A_1'}{\partial \theta} \label{eq:rho1-pol}\\
P_1'&=&\gamma \rho_1'\ . \label{eq:P1-pol}
\end{eqnarray}
We solve the G-S equation (\ref{eq:GS-pol}) numerically in the region $r<1$, $0<\theta<\pi$ with boundary conditions on $r=1$ and $\theta=0,\ \pi$ (see next subsection).

\subsection{Boundary conditions}
At FRWF ($r=1$), deviation of any quantity from the initial equilibrium state must vanish (see equations [\ref{eq:B1}]-[\ref{eq:P1}] and [\ref{eq:v1x-pol}]-[\ref{eq:P1-pol}]). Thus, $A_1'=const.$, and should be a $C^1$-class continuous function at $r=1$. Without any loss of generality, we can set $A_1'=0$ and $\partial_r A_1'=0$ on $r=1$. 

Note that if we impose a Dirichlet type boundary condition $A_1'=0$ on $r=1$, another condition $\partial_r A_1'=0$ (this means $\mbox{\boldmath $B$}_1'=0$ on $r=1$) is automatically satisfied. This is easily understood from equation (\ref{eq:GS-pol}). The first two terms in the left-hand side vanish on $r=1$, then $\partial_r A_1'$ must vanishes if these quantities are continuous at this point. 

The boundary condition at $y=0$ is not trivial as at $r=1$. This boundary corresponds to the slow shock, and is the interface between the inflow region and the shocked region (reconnection jet). We need precise physical discussion to determine the junction condition for this boundary. Instead of solving this difficult problem, we set the boundary condition at this interface by adopting the result of our numerical simulation (Nitta et al. 2001). 

Our previous simulation gives information for every quantity just outside the slow shock. We adopt the following simple functional form of $A_1'$ approximating the boundary values of $A_1'$ from our previous simulation, 
\begin{eqnarray}
&&A_1'(|x|, y=0) \nonumber \\
&=&(-\frac{|x|}{x_c}+1) A_0 \\
&&(\mbox{for} \ |x| \leq x_m) \nonumber \\
&=&\frac{x_m/x_c-1}{1-x_m}(|x|-1) A_0\\
&&(\mbox{for} \ x_m<|x|<1) \nonumber \ .
\end{eqnarray}
where $A_0 \ (\ll 1)$ is the value of $A_1'$ at the reconnection point $x=y=0$ (this shows the reconnection rate $v_{1y}'$ at $x=y=0$: see equation [\ref{eq:v1y}]), $x_c$ is the location of the contact discontinuity (the interface between the original current sheet plasma and the reconnected inflow plasma), and $x_m$ is the location of the minimum of $A_1'$. Thus, we have set a complete Dirichlet type boundary problem for the linearized G-S equation (\ref{eq:GS}). We set $A_0$, $x_c$ and $x_m$ to be consistent with our numerical simulation. For example, the figures shown in the next section are for the case $A_0=0.055$, $x_c=0.64$, $x_m=0.84$. 

We compare the model boundary-value defined above and the simulation result (Case A of Nitta et al. 2001) of $A_1'$ along the slow shock in figure \ref{fig:bc}. The solid and dotted lines show the approximated model adopted here and the simulation result, respectively. The actual slow shock has a small tilt angle from the $x$-axis, so that, when plotting the simulation result (dotted line), we have projected the $A_1'$ values obtained at the slow shock onto the $x$-axis. The two straight line-segments constituting the model boundary condition have been determined in such a way that they pass the simulated values at the X-point ($x=0, \ A_1'=0.055$: this value represents the reconnection rate), at the contact point ($x=0.64, \ A_1'=0$: the vanishing $A_1'$ corresponds to the contact point between the initial current sheet plasma and the inflow plasma), and at the minimum $A_1'$ ($x=0.84$). We can see that the solid line approximates the dotted line reasonably well. 

\placefigure{fig:bc}

\section{Self-similar solutions}
By using the SOR routine, we numerically solve the linearized G-S equation (\ref{eq:GS-pol}) under the Dirichlet type boundary condition as discussed in the previous section. A typical result ($A_0=0.055$, $x_c=0.64$, $x_m=0.84$, see the last section) is obtained in the figures \ref{fig:A1} ($A_1'$), \ref{fig:v1} ($v_{1y}'$), and \ref{fig:rho1} ($\rho_1'$). 

\placefigure{fig:A1}
\placefigure{fig:v1}
\placefigure{fig:rho1}

Let us compare the results of this semi-analytic study with the result of our simulation in Nitta et al. 2001. Figures \ref{fig:A1}, \ref{fig:v1} and \ref{fig:rho1} are the semi-analytic results showing the first order quantities solved here: magnetic vector potential ($z$-component), velocity ($y$-component), and mass density, respectively. Figures \ref{fig:A1-s}, \ref{fig:v1-s} and \ref{fig:rho1-s} show the numerical result in Nitta et al. 2001: the variation from the initial equilibrium of magnetic vector potential ($z$-component), velocity ($y$-component), and mass density, respectively. Each quantity is renormalized in the way explained in section 2.2 (see [\ref{eq:n-r}]-[\ref{eq:n-P}]). Small numbers affixed to some of the contours denote the level values at these points. We can easily find that the result of the semi-analytic study is consistent with the result of our numerical simulation not only in the topological sense of contours, but also quantitatively. 

\placefigure{fig:A1-s}
\placefigure{fig:v1-s}
\placefigure{fig:rho1-s}

We have here succeeded to verify that 1) our result of numerical simulation (Nitta et al. 2001) truly represents a self-similar growth, 2) this self-similar growth is stable over a long duration. The spatial dynamic range of the numerical study is limited (at most $\sim 10^3$) by technical restriction ( e.g., restriction of memory and CPU time), but the fact that the result of numerical simulation is consistent with the semi-analytic solution suggests that the self-similar growth will continue indefinitely. On the other hand, the stability of semi-analytic solution with respect to any MHD mode is certified by numerical simulation. We can say that our semi-analytic study and numerical simulation are complementary to each other.

\section{Summary \& Discussion}

\subsection{Summary}

Together with our time-dependent numerical simulation (see Nitta et al. 2001), the linearized perturbation solution discussed in this paper has ensured the existence of self-similar growth of fast magnetic reconnection. 

Thus we propose here the new model describing ``self-similar evolution of fast reconnection''. The time dependent simulation directly solving the MHD equations numerically is effective to check the stability of the evolving system. However, the duration of simulation is restricted owing to the restriction of computer memory and run time, or stability of the simulation code itself. Hence, even if we find the self-similar (-like) behavior in the result of simulation, we cannot be convinced that this behavior is a true one that may continue indefinitely. 

On the other hand, an analytical study has the following properties. If one solves the MHD equation in the zoom-out coordinate under the stationary assumption and obtain a (semi-) analytic solution which is identical to the solution of numerical simulation, we can be convinced that the numerically obtained self-similar-like solution is truly the self-similar one. On the other hand, an analytical study of the stationary equation does not give information on the stability of the solution. 

From these arguments, we can reach the following conclusion. Our time-dependent numerical study and analytical study are complimentary to each other in establishing the model of ``self-similar evolution of fast reconnection''.

\subsection{Consistency between semi-analytic and numerical studies}
In section 4, we compared our semi-analytic result with numerical result. We can notice significant similarity between semi-analytic and numerical results. 

If we compare them quantitatively in detail, we can notice that these two results are consistent in the region in which $v_{1y}'<0$. However, in the region in which $v_{1y}'>0$, the value of semi-analytic result is somewhat different from the numerical result. In this region, strong fast-mode compression (``piston effect'' of the reconnection jet) takes place, and, strictly speaking, our linearized treatment might not be suitable to be applied here.

Thus, we can conclude that the inflow region is well described by our linearized theory. This is obviously due to the very small reconnection rate ($\sim 10^{-2}$, see Nitta et al. 2001). However, the mechanism leading to such a small reconnection rate is still unclear. In this self-similar reconnection model, the reconnection rate should be self consistently determined by the dynamical evolution process, and this point will be discussed in our forthcoming paper.

\subsection{Condition for self-similar reconnection}
\label{sec:cond}

The above type of evolutionary reconnection will be realized in the systems in which 1) the initial spatial scale of the disturbance ( e.g., scale of microscopic instabilities which will lead to the anomalous transport phenomena $\sim 10^0$ [m] if we estimate it by ion Larmor radius in the typical case of solar flares) is much smaller than the entire spatial scale of the system ( e.g., the curvature radius of the loop of magnetic flux tube $\sim 10^7-10^8$ [m]  in the typical case of solar flares), and 2) there is no proper spatial scale except the one which expands as time proceeds (e.g., the radius of FRWF in our case). In such a system, the magnetic reconnection triggered by the initial disturbance can evolve as the FRW propagates (see section 2 of Nitta et al. 2001). The spatial scale of FRW propagation is the only proper scale of the system if it is much larger than the initial current sheet thickness. The spatial dynamic range of the evolution is determined by the ratio of the entire system scale to the initial disturbance scale. In the above mentioned case of typical solar flares, it will be of the order of $10^7$. In such a system, the external boundary does not affect the evolution for a certain amount of time after the onset of reconnection. This means that the system can freely evolve independent of the outer boundary condition. Such a very wide dynamic range and only one evolving spatial scale strongly suggest the possibility of self-similar solution. In fact, we have obtained the self-similar solution both numerically and analytically.

\subsection{Near FRWF}
The region near the FRWF ($r \sim 1$) actually has a complicated nature. The region near the spearhead of reconnection jet (roughly to say, $r \sim 1, \ 0<\theta<\pi/8$) is influenced  not only by the fast-mode rarefaction wave, but also by the fast-mode compression wave induced by the ``piston effect'' of the jet. Hence one might think that the word ``FRWF'' (fast-mode {\it rarefaction} wave front) is somewhat misleading.

However, we should notice that, in our linearized treatment, any non-linear interaction between waves is completely neglected. Thus the fast-mode rarefaction wave emitted from the vicinity of reconnection point keeps a circular shape which is truly located at $r=1$ in the zoom-out coordinate even if the compressional mode is superimposed. From this reason, we may continue to call it ``FRWF'' as we used this term in our previous paper.

\subsection{Relation to reconnection jet}

We have treated only the inflow region in this paper in spite of the importance of the reconnection jet. This is owing to the mathematical difficulty to treat the reconnection jet where deviation from the initial state is very large and the linear analysis we have used here breaks down. The complete solution which covers the entire region has not been obtained yet. 

However, we should note the importance to study the property of the inflow region. As Priest \& Forbes (1986) discussed, the property of the inflow region crucially shows the relation to previous classical models of reconnection. We can see that our new model is an extension of the Petschek model to the evolutionary model in free space (see section \ref{sec:Com}). The fast-mode rarefaction dominated inflow restricts the reconnection rate as in the Petschek model (see Priest \& Forbes 1986 or Vasyliunas 1975). In our case, the region near the X-point is filled with the fast-mode rarefaction dominated inflow. Hence we may expect that our model has a maximum value of the reconnection rate. This point will be clarified by our forthcoming study of the reconnection jet which induces the fast-mode rarefaction wave. 

In our analytic work, the boundary condition at $y=0$, which corresponds to the junction condition to the reconnection jet, is artificially imposed approximating the result of our numerical simulation. Needless to say, this boundary condition is more important than other boundary conditions (e.g., conditions on $r=1$ or $\theta=\pi/2$) because it represents physical information about the reconnection jet, and it crucially influences the solution of inflow region. 

One might think that the similarity of the boundary condition at $y=0$ is not trivial, and the above boundary condition is imposed {\it ad hoc}. However, we must note that everything evolves self-similarly in this situation. In the self-similar stage $V_{A0} t \gg D$, the entire reconnection system including reconnection jet has no proper length other than the scale of FRWF $V_{A0} t$. Thus, it naturally leads that the reconnection jet itself will grow self-similarly, and hence, the similarity growth of reconnection jet is realized.

\subsection{Relation to diffusion region}

The self-similar stage realizes after the system size $V_{A0} t$ sufficiently exceeds any finite fixed spatial scale, i.e., the initial thickness of the current sheet and size of the diffusion region. Since both of these scales are estimated as $D$, the self-similar stage can be realized for $V_{A0} t \gg D$ as noted in section \ref{sec:intro}. There is no need to go into details about the structures of them because any fixed scale is negligible in the self-similar stage. In other words, this shows the universality of the self-similar evolution. The self-similar evolution does not depend on detailed structure of the current sheet. We can adopt any current sheet model if it satisfies the initial MHD equilibrium. The self-similar evolution also does not depend on detailed resistivity model and structure of the diffusion region. The only function of the diffusion region which we need is to reconnect the field lines in a desired speed. We may expect that this is possible by dynamical change of the thickness of diffusion region. In fact, our numerical simulation showed that our new model is insensitive to the resistivity model (this is an important property of the fast reconnection).

\subsection{Comparison with previous steady and unsteady models}
\label{sec:Com}

There are several theoretical models for steady state and time-depending magnetic reconnection. We compare our self-similar evolution model with these previous models. Our discussion is focused only on fast reconnection, because very quick energy conversion frequently observed in astrophysical phenomena suggests that fast reconnection should be considered as the responsible mechanism. 

The Petschek model (Petschek 1964) is characterized by a pair of slow shocks and a fast-mode rarefaction wave in the inflow region. This fast-mode rarefaction wave propagates outward from the reconnection point. As a result of the fast-mode rarefaction, the gradient of the magnetic field strength near the neutral point decreases due to the bending of magnetic field lines (Vasyliunas 1975). This process suppresses and limits the diffusion speed, and hence the reconnection rate. 

The Sonnerup model (Sonnerup 1970) is developed from the Petschek model. This model is characterized by a pair of slow shocks and the hybrid of fast-mode and slow-mode rarefaction. The fast-mode rarefaction wave is produced at the central region as in the Petschek model, but the slow-mode rarefaction wave is injected from the boundary. Because of the hybrid nature of rarefaction modes, the gradient of the magnetic field strength near the neutral point does not decrease as in the Petschek model. Therefore the Sonnerup model attains the maximum reconnection rate possible for magnetic energy converters (Priest \& Forbes 1986). A modified model of the Sonnerup solution was presented by Priest \& Forbes (1986).  They found a solution with a diffuse slow-mode rarefaction waves spread throughout the inflow region, while in the original Sonnerup model the rarefaction waves were treated as discontinuities.

The property of the central region of our model can be categorized into the ``almost-uniform reconnection'' which means that magnetic field lines in the inflow region are almost straight and uniform (see Priest \& Forbes 2000). Priest \& Forbes (1986) presented a unification scheme of 2-dimensional steady almost-uniform reconnection in a finite space filled with incompressible plasma. They found a family of continuous solutions characterized by a non-dimensional parameter ``$b_0$''. The Petschek ($b_0=0$) and Sonnerup ($b_0=1$) types are particular cases. Their result includes other types, e.g., Sweet-Parker type, flux-pile-up type and stagnation-flow type according to the value of $b_0$. These solutions ($b_0>0$) other than the Petschek type are characterized by the hybrid of slow-mode and fast-mode rarefaction. Our self-similar evolution is never influenced by the outer conditions, so the hybrid rarefaction never takes place in our case because there are no boundary-imposed waves in our case. We can conclude that the unique candidate which is worth comparing with the central region of our model is the Petschek model. 

Figures \ref{fig:A1} and \ref{fig:rho1} clearly show decreases of the magnetic field and gas pressure, thus the fast-mode rarefaction dominates in the vicinity of the diffusion region. Obviously the central region of this self-similarly evolving system is of the Petschek-type. Therefore, the reconnection rate might be limited in a way similar to the original Petschek model. This point will be clarified in our forthcoming paper. 

Far from the central region, there is a region in which the fast-mode compression takes place by the ``piston effect'' of the reconnection jet. This fast-mode compression causes the vortex-like return flow (see figure 13 of Nitta et al. 2001). The combination of the inner structure which resembles the original Petschek model and the outer structure involving the flow vortices characterizes the evolutionary process, and the entire system unlimitedly expands self-similarly. 

Biernat, Heyn and Semenov (1987) and Semenov et al. (1992) analytically studied the evolutionary process of fast magnetic reconnection in a similar situation to our case. Their analysis gives a general formalism of spontaneous time-varying reconnection. They succeeded to obtain the solution for Petschek-type reconnection. However, in their works, plasma in the inflow region is assumed to be incompressible for analytical convenience. This means that the sound speed is infinitely large even when the Alfv\'{e}n speed is finite. This is equivalent to assuming that the inflow region is filled with extremely high $\beta$ plasmas (note that $\beta \sim$ (sound speed)$^2$/ (Alfv\'{e}n speed)$^2$). Needless to say, this assumption is unsuitable for most cases of astrophysical problems. Contrary to our case, fast-mode waves emitted from the central region instantly propagates to infinity. Although the proper spatial scale determined by the fast-mode wave propagation does not exist in their case, there is a possibility that the structure formed by slow-mode wave can be self similar, because the propagation speed of the slow-mode wave is finite in their case. Our work is understood as an extension of their works to realistic astrophysical systems filled with low $\beta$ compressible plasma.

\subsection{Astrophysical applications}
In the present level of our observational technology, only one object in which we can obtain enough spatial and time resolution of the distributions of plasma parameters is solar flares. We expect the application of our self-similar reconnection model to the solar flares. 

\subsubsection{Dimming and inflow structure}
The propagation speed of FRW is estimated to be $\sim 10^6$ [m s$^{-1}$] for solar corona. Hence, the duration of the self-similar evolution is typically $10^1-10^2$[s] for the flux tubes having the spatial dimension $\sim 10^7-10^8$ [m]. The evolution having such a time scale is able to be resolved by the {\it Solar-B} project (required cadence of {\it Solar-B} X-ray telescope [XRT] is 2 sec., which is sufficient to resolve the expected time evolution of the self-similar reconnection, see Golub 2000.). Especially we expect that the ``dimming'' will be detected by {\it Solar-B}. Expected X-ray image of the dimming is argued in Nitta et al. 2001 (see section 4.5). We consider that the dimming is a naturally expected phenomenon at the inflow region of our self-similar reconnection model. If we obtain the information of velocity field and other plasma parameters at the reconnection site by observation, we can inspect our model more directly. 

The {\it Solar-B} satellite will clarify also the velocity field, plasma density and temperature near the reconnection point by the extreme ultraviolet imaging spectrometer (EIS). The combination of XRT and EIS will reveal a detailed structure of the reconnection system in solar flares. Yokoyama et al. (2001) found the evidence for reconnection inflow in a flare on 1999 March 18. The estimated inflow speed is 5 km s$^{-1}$ which corresponds to the reconnection rate 0.001-0.03 (this result is not far from the expected value 0.05 from our model). Their result encourages our theoretical effort of applying our self-similar model to the inflow region of real solar flares. We hope that the expected feature of the inflow region by virtue of our self-similar reconnection model will be inspected by the {\it Solar-B} project.

\subsubsection{Application to realistic cases}
Roussev et al. (2001 a, b, c) predicted emission from explosive events in the solar transition region induced by magnetic reconnection. They performed MHD numerical simulations including thermal conduction, radiative losses and volumetric heating. They treated spontaneous evolution of magnetic reconnection with the initial conditions similar to our model (anti-parallel magnetic field and asymptotically uniform plasma distribution) and with the resistivity (artificially localized and kept as a constant) that is also very similar to ours.  The resultant dynamical properties are similar to our study, indicating that the essence of our model is applicable to more complicated situations with more realistic thermal and radiative properties.

\subsection{Expectation of 3D self-similar reconnection}
Our present discussion is restricted to 2D reconnection in this as well as in the previous paper (Nitta et al. 2001) for simplicity. However, actual current sheet systems have finite depth in the direction perpendicular to the figures of this paper (for example, radius of the flux tube or length of the arcade structure of a bunch of flux tubes). We believe the essence of the self-similar growth has already been understood by our 2D approximation. When we discuss the phenomenology of the evolutionary reconnection by virtue of the self-similar reconnection model in near future, we need a more precise 3D model. 

We can expect that the self-similar growth also takes place in a 3D system. As discussed in section \ref{sec:cond}, a sufficiently evolved system has only one proper spatial scale, i.e., the scale of FRWF $V_{A0} t$ which is increasing in proportion to time (note that the propagation speed of the fast-mode is almost isotropic in a 3D current sheet system filled with a low $\beta$ plasma, hence the shape of FRWF will be spherical). This is just the situation that leads to realize the self-similar expansion. Of course, detailed structure of the 3D reconnection system will be different from our 2D model, but essential properties will be common between them.

\acknowledgments

One of the author SN is grateful to T. Kudoh at National Astronomical Observatory of Japan for useful comments and fruitful discussion.

\appendix
\section{Appendix A: Derivation of MHD equations in the Zoom-out coordinate}

We show the derivation of equations (\ref{eq:cont-z})-(\ref{eq:pol-z}) briefly in this appendix. The zoom-out coordinate is defined by equation (\ref{eq:n-r}). The velocity in this new coordinate has been given in equation (\ref{eq:n-v}). This equation is deformed to
\begin{equation}
V_{A0} \mbox{\boldmath $v$}'=\mbox{\boldmath $v$}-V_{A0} \mbox{\boldmath $r$}' \ .
\end{equation}
The second term of the right hand side is the apparent converging flow properly appearing in the zoom-out coordinate. Physical quantities are put into non-dimensional forms as in (\ref{eq:n-v})-(\ref{eq:n-P}). With these preparations, we can derive the modified MHD equations in the zoom-out coordinate. 

Any scalar field, e.g., $f(\mbox{\boldmath$r$}, t)$ is transformed in the zoom-out coordinate as $f(\mbox{\boldmath$r$}'[\mbox{\boldmath$r$}, t], t)$. Thus the time derivative of any scalar field can be transformed as
\begin{eqnarray}
\left. \frac{\partial f}{\partial t}\right|_{\mbox{\boldmath$r$}}&=&\left. \frac{\partial f}{\partial t}\right|_{\mbox{\boldmath$r$}'}+\frac{\partial \mbox{\boldmath$r$}'}{\partial t} \cdot \mbox{\boldmath $\nabla$}' f \\
&=&\left. \frac{\partial f}{\partial t}\right|_{\mbox{\boldmath$r$}'}-\frac{1}{t} \mbox{\boldmath$r$}' \cdot \mbox{\boldmath $\nabla$}' f
\end{eqnarray}
where subscripts $\mbox{\boldmath$r$}$ and $\mbox{\boldmath$r$}'$ show the time derivative in the fixed coordinate and zoom-out coordinate, respectively, and $\mbox{\boldmath $\nabla$}'$ denotes the spatial derivative with respect to $\mbox{\boldmath$r$}'$. Here we have used (\ref{eq:n-r}) to derive the second equality. The transform of time derivative for any vector field is similar to this. 

The gradient of any scalar field is transformed as
\begin{equation}
\nabla f|_{\mbox{\boldmath$r$}}=\frac{1}{V_{A0} t} \mbox{\boldmath $\nabla$}' f\ .
\end{equation}

The divergence and rotation of any vector field, e.g., $\mbox{\boldmath$A$}$ are transformed as
\begin{equation}
\nabla \cdot \mbox{\boldmath$A$}=\frac{1}{V_{A0} t} \mbox{\boldmath $\nabla$}' \cdot \mbox{\boldmath$A$}
\end{equation}
and
\begin{equation}
\nabla \times \mbox{\boldmath$A$}=\frac{1}{V_{A0} t} \mbox{\boldmath $\nabla$}' \times \mbox{\boldmath$A$} \ ,
\end{equation}
respectively. By using these transformed notations, we can rewrite the MHD equations as follows. 
\begin{equation}
\frac{\partial \rho}{\partial t}+\nabla \cdot (\rho \mbox{\boldmath$v$})=0 \ \ (\mbox{continuity eq.})
\end{equation}
\begin{equation}
\frac{\partial (\rho \mbox{\boldmath$v$})}{\partial t}+\nabla \cdot \left(\rho \mbox{\boldmath$v$} \mbox{\boldmath$v$}-\frac{1}{\mu_0} \mbox{\boldmath$B$} \mbox{\boldmath$B$} \right)+\nabla \left(P+\frac{\mbox{\boldmath$B$}^2}{2 \mu_0} \right)=0 \ \ (\mbox{momentun eq.})
\end{equation}
\begin{equation}
\frac{\partial \mbox{\boldmath$B$}}{\partial t}+\nabla \cdot (\mbox{\boldmath$v$} \mbox{\boldmath$B$}-\mbox{\boldmath$B$} \mbox{\boldmath$v$})=0 \ \ (\mbox{induction eq.})
\end{equation}
\begin{equation}
P \rho^{-\gamma}=Const. \ \ (\mbox{polytrope relation})
\end{equation}
As a representative case, we will demonstrate the derivation of the modified continuity equation. The derivations for other modified MHD equations are quite similar to this. 

The original continuity equation
\begin{equation}
\left.\frac{\partial \rho}{\partial t}\right|_{\mbox{\boldmath$r$}}+\nabla \cdot (\rho \mbox{\boldmath$v$})=0
\end{equation}
is transformed to 
\begin{equation}
\left.\frac{\partial \rho}{\partial t}\right|_{\mbox{\boldmath$r$}'}-\frac{\mbox{\boldmath$r$}'}{t} \cdot \mbox{\boldmath $\nabla$}' \rho+\frac{1}{V_{A0} t} \mbox{\boldmath $\nabla$}' \cdot (\rho \mbox{\boldmath$v$})=0\ .
\end{equation}
Here we suppose the stationariness in the zoom-out coordinate (self-similar expansion). It makes the first term vanish. By substituting the normalization relations (\ref{eq:n-v})-(\ref{eq:n-P}) into this transformed equation, we easily obtain the following equation, 
\begin{equation}
\mbox{\boldmath$r$}' \cdot \mbox{\boldmath $\nabla$}' \rho'-\mbox{\boldmath $\nabla$}' \cdot [\rho'(\mbox{\boldmath$v$}'+\mbox{\boldmath$r$}')]=0 \ .
\end{equation} 
This equation is expanded to the form
\begin{equation}
-\mbox{\boldmath $\nabla$}' \rho' \cdot \mbox{\boldmath$v$}'-\rho' \mbox{\boldmath $\nabla$}' \cdot \mbox{\boldmath$v$}'-\rho' \nabla' \cdot \mbox{\boldmath$r$}'=0 \ .
\end{equation}
We combine the first two terms to obtain $-\mbox{\boldmath $\nabla$}' \cdot (\rho' \mbox{\boldmath$v$}')$. We should note that in the two-dimensional problem,
$\mbox{\boldmath $\nabla$}' \cdot \mbox{\boldmath$r$}'=2$. Thus we finally obtain the transformed continuity equation (\ref {eq:cont-z}). 

Similarly we can easily obtain the other modified equations (\ref{eq:mom-z})-(\ref{eq:pol-z}). Each derivation is not difficult, but is very tedious, hence we omit detailed derivation here.

\section{Appendix B: Perturbative relations}
By assuming that the deviation by reconnection from the initial equilibrium state is very small, we derive the linearized relations. We approximately expand each quantity as in equations (\ref{eq:B0+1})-(\ref{eq:rho0+1}) with (\ref{eq:b0})-(\ref{eq:rho0}), and substitute them into the modified MHD equations (\ref{eq:cont-z})-(\ref{eq:pol-z}). We can obtain the linearized MHD equations as is demonstrated below. 

For example, we will demonstrate the derivation of the linearized continuity equation (\ref{eq:lin-con}). The derivation of other linearized MHD equations is quite similar. We substitute the expansion forms (\ref{eq:rho0+1}) and (\ref{eq:v0+1}) into the modified continuity equation (\ref{eq:cont-z}). By expanding equation (\ref{eq:cont-z}) up to the first order of magnitude, we obtain two equations: the zeroth order equation 
\begin{equation}
\mbox{\boldmath $\nabla$}' \cdot (\rho_0' \mbox{\boldmath$v_0$}')=-2\rho'_0
\end{equation}
and the first order equation
\begin{equation}
\mbox{\boldmath $\nabla$}' \cdot (\rho'_0 \mbox{\boldmath$v_1$}'+\rho'_1 \mbox{\boldmath$v_0$}')=-2 \rho'_1\ .
\end{equation}
The left hand side of the zeroth order equation is reduced to
\begin{eqnarray}
\mbox{\boldmath $\nabla$}' \cdot (\rho_0' \mbox{\boldmath$v_0$}')&=&\mbox{\boldmath $\nabla$}' \cdot (-\mbox{\boldmath$r$}')\\
&=&-2 \ .
\end{eqnarray}
The right hand side of this equation is -2. Thus, we can see that the zeroth order equation is trivial and has no information. The left hand side of the first order equation is reduced to
\begin{eqnarray}
\mbox{\boldmath $\nabla$}' \cdot (\rho'_0 \mbox{\boldmath$v_1$}'+\rho'_1 \mbox{\boldmath$v_0$}')&=&\mbox{\boldmath $\nabla$}' \cdot \mbox{\boldmath$v_1$}'+\mbox{\boldmath $\nabla$}' \cdot (-\rho'_1 \mbox{\boldmath$r$}')\\
&=&\mbox{\boldmath $\nabla$}' \cdot \mbox{\boldmath$v_1$}'-\mbox{\boldmath$r$}' \cdot \mbox{\boldmath $\nabla$}' \rho'_1-\rho'_1 \mbox{\boldmath $\nabla$}' \cdot \mbox{\boldmath$r$}'\ .
\end{eqnarray}
By noting that the last term is $-2 \rho'_1$ and together with the right hand side, we obtain the linearized continuity equation (\ref{eq:lin-con}). By similar reduction, we obtain other linearized equations (\ref{eq:lin-mom})-(\ref{eq:lin-pol}).

\section{Appendix C: Derivation of the G-S equation and other relations}
We here show the reduction of (\ref{eq:v1x})-(\ref{eq:P1}). Let us start with the deformation of the linearized momentum equation (\ref{eq:lin-mom}). Each term of that equation is reduced as following: 
\begin{equation}
-\mbox{\boldmath $i$} \cdot \mbox{\boldmath $\nabla$}' \mbox{\boldmath $B$}_1'=-\frac{\partial^2 A_1'}{\partial x \partial y} \mbox{\boldmath $i$}+\frac{\partial^2 A_1'}{\partial x^2} \mbox{\boldmath $j$}
\end{equation}
\begin{equation}
\mbox{\boldmath $\nabla$}' (\mbox{\boldmath $i$} \cdot \mbox{\boldmath $B$}_1')=\frac{\partial^2 A_1'}{\partial x \partial y} \mbox{\boldmath $i$}+\frac{\partial^2 A_1'}{\partial y^2} \mbox{\boldmath $j$}
\end{equation}
\begin{equation}
-\mbox{\boldmath $r$}' \cdot \mbox{\boldmath $\nabla$}' \mbox{\boldmath $v$}_1'=-\left[x \frac{\partial v'_{1x}}{\partial x}+y \frac{\partial v'_{1x}}{\partial y}\right]\mbox{\boldmath $i$}-\left[x \frac{\partial v'_{1y}}{\partial x}+y \frac{\partial v'_{1y}}{\partial y}\right]\mbox{\boldmath $j$}\ .
\end{equation}
Then we obtain the reduced form of each component of the equation (\ref{eq:lin-mom}) as
\begin{equation}
x \frac{\partial v'_{1x}}{\partial x}+y \frac{\partial v'_{1x}}{\partial y}=0 \ ,\label{eq:momx}
\end{equation}
\begin{equation}
-\left[x \frac{\partial v'_{1y}}{\partial x}+y \frac{\partial v'_{1y}}{\partial y}\right]+\left(\frac{\partial^2}{\partial x^2}+\frac{\partial^2}{\partial y^2}\right) A_1'=0 \ . \label{eq:momy}
\end{equation}

Similarly, from the linearized induction equation (\ref{eq:lin-ind}), we obtain the following two equations: 
\begin{equation}
x \frac{\partial^2 A_1'}{\partial x \partial y}+y \frac{\partial^2 A_1'}{\partial y^2}-\frac{\partial v'_{1y}}{\partial y}=0 \ ,\label{eq:indx}
\end{equation}
\begin{equation}
x \frac{\partial^2 A_1'}{\partial x^2}+y \frac{\partial^2 A_1'}{\partial y \partial x}-\frac{\partial v'_{1y}}{\partial x}=0 \ . \label{eq:indy}
\end{equation}
We should note that these two equations from the induction equation have the same information in common
\begin{equation}
x \frac{\partial A_1'}{\partial x}+y \frac{\partial A_1'}{\partial y}-A_1'-v'_{1y}=(\mbox{Const. independent of } x,\ y) \ .
\end{equation}
The vector potential $A_1'$ has a freedom of an indeterminate constant, then the constant on the right hand side can be absorbed into the potential, and we obtain equation (\ref{eq:v1y}). 

We substitute the two components of the induction equation (\ref{eq:indx}) and (\ref{eq:indy}) into (\ref{eq:momy}), and obtain a single partial differential equation (\ref{eq:GS}) for $A_1'$ after simple calculations. This is the G-S equation. 

Another component of the momentum equation (\ref{eq:momx}) is transformed to a simple form: 
\begin{equation}
\mbox{\boldmath $r$}' \cdot \mbox{\boldmath $\nabla$}' v'_{1x}=0 \ .
\end{equation}
This equation clearly shows that $v'_{1x}$ depends only on the angle $\theta$ from the $x$-axis. We should remember that $v'_{1x}=0$ on $r'=1$ (FRWF). Thus this equation implies $v'_{1x}=0$ everywhere in our linearized analysis. This is the equation (\ref{eq:v1x}). 

The linearized continuity equation (\ref{eq:lin-con}) is deformed by using the above results. The first term is reduced by the help of (\ref{eq:indx}) as 
\begin{eqnarray}
\mbox{\boldmath $\nabla$}' \cdot \mbox{\boldmath $v$}_1'&=&\frac{\partial v_{1y}'}{\partial y}\\
&=&x \frac{\partial^2 A_1'}{\partial x \partial y}+y \frac{\partial^2 A_1'}{\partial y^2}\\
&=&\mbox{\boldmath $r$}' \cdot \mbox{\boldmath $\nabla$}' \frac{\partial A_1'}{\partial y} \ .
\end{eqnarray}
Together with the second term of (\ref{eq:lin-con}), we can see that the continuity equation implies 
\begin{equation}
\mbox{\boldmath $r$}' \cdot \mbox{\boldmath $\nabla$}' (\frac{\partial A_1'}{\partial y}-\rho'_1)=0 \ .
\end{equation}
Similarly to the discussion of the previous paragraph, we obtain the result (\ref{eq:rho1}) because we have set the boundary condition $\partial A_1'/\partial y=0$ and $\rho'_1=0$ on FRWF. The relation (\ref{eq:P1}) is easily obtained from (\ref{eq:lin-pol}). 

\appendix

\figcaption[bc.eps]{
Boundary condition of $A_1'$ along the slow shock ($y=0$). The dotted line shows the result of our simulation in Nitta et al. 2001 (Case A). The solid line is the approximated boundary condition adopted in this work. We can see the approximated model is very simple and gives good agreement with the simulation result. 
\label{fig:bc}
}

\figcaption[A1.eps]{
Contours of the perturbed flux function ($z$-component of magnetic vector potential) $A_1'$. The solid (dashed) contours correspond to positive (negative) value of $A_1'$ (The same applies to figures \ref{fig:v1}, \ref{fig:rho1}, \ref{fig:A1-s}, \ref{fig:v1-s} and \ref{fig:rho1-s}). The contact discontinuity is located at $x=x_c \sim 0.64, \ y=0$. The location of the maximum value is at the reconnection point ($x=y=0$), and that of minimum value is near the spearhead of reconnection jet ($x=x_m \sim 0.84$, \ $y=0$). Hence the first order magnetic field is turned counter-clockwise along the left-half arcs, and clockwise in the right-half arcs. The zeroth order magnetic field is directed to the right in the upper-half plane. Thus the strength of the magnetic field is reduced by the first order variation in the vicinity of the reconnection point. This region is filled with the fast-mode rarefaction dominated inflow. This property is common to the original Petschek model. On the contrary, the strength of the magnetic field is enhanced by the first order variation near the spearhead of reconnection jet. This is caused by the fast-mode compression (piston effect) produced by the reconnection jet. This compressed region is characteristic to our dynamically evolving system, and quite different from previous stationary models. 
\label{fig:A1}
}

\figcaption[A1-s.eps]{
The variation of flux function from the initial state (result of our simulation in Nitta et al. [2001]). We can see essentially the same feature in figures \ref{fig:v1-s} and \ref{fig:rho1-s} except that the configuration near the $x$-axis ($y < 0.05$) is rather different. This is due to the existence of reconnection jets which are neglected in our present analytical study. 
\label{fig:A1-s}
}

\figcaption[v1.eps]{
Contours of the $y$-component $v_{1y}'$ of the velocity perturbation. The $x$-component of the velocity perturbation is zero in the linear analysis. We should note that the value of $v_{1y}'$ is equivalent to the strength of electric field in the $z$ direction (so-called reconnection electric field) in the normalized unit. 
The value near the reconnection point is negative. This means that the inflow is sucked into the reconnection jet. Positive values exist near the right-half outer boundary. This is the outflow gushing out from the region near the spearhead of the reconnection jet. Hence, this outflow is a secondary flow properly characteristic to the evolutionary process. This outflow enhances the vortex-like flow. The value at the reconnection point ($x=y=0$) represents the reconnection rate of the system, and the value is 0.055 from our simulation (Nitta et al. 2001). 
\label{fig:v1}
}

\figcaption[v1-s.eps]{
The inflow velocity ($y$-component) (result of our simulation if Nitta et al. [2001]). 
\label{fig:v1-s}
}

\figcaption[rho1.eps]{
Contours of the perturbed density $\rho_1'$. If we divide this value by  adiabatic constant $\gamma$, these contours show the perturbation in gas pressure $P_1'$ in the normalized unit. The density (hence pressure) is reduced near the reconnection point, and enhanced near the spearhead of the reconnection jet. These features are also caused by the fast-mode rarefaction and compression as discussed in the caption for figure \ref{fig:A1}. 
\label{fig:rho1}
}

\figcaption[rho1-s.eps]{
The variation of density distribution from the initial state (result of our simulation in Nitta et al. [2001]). 
\label{fig:rho1-s}
}


\begin{thebibliography}{99}
\baselineskip=1.0pc

\bibitem[Biernat 1987]{BHS}
Biernat,H.K., Heyn,M. \& Semenov,V.S., 1987, JGR, 92, A4, 3392

\bibitem[Golub 2000]{SB}
Golub,L., 2000, "XRT Science Objectives in The 2nd Solar-B Science Meeting", P.47

\bibitem[Nitta et al. 2001]{Nit01}
Nitta, S., Tanuma, S., Shibata, S., Maezawa, K., 2001, ApJ, 550, 1119

\bibitem[Parker 1963]{Par}
Parker, E. N. 1963, ApJ Suppl. Ser., 8,177

\bibitem[Petschek 1964]{Pet}
Petschek,H.E. 1964, NASA Spec. Publ., 50, AAS-NASA Symposium on Physics of Solar Flares, 425

\bibitem[Priest \& Forbes 1986]{P-F}
Priest,E.R. \& Forbes,T.G. 1986, J.Geophys.Res, 91, 5579

\bibitem[Priest \& Forbes 2000]{P-F-book}
Priest,E.R. \& Forbes,T.G. 2000, Magnetic reconnection (Cambridge University Press)

\bibitem[Roussev 1]{R1}
Roussev,I., Galsgaard,K., Erd\'{e}lyi,R. \& Doyle,J.G., 2001a, A\&A, 370, 298

\bibitem[Roussev 1]{R2}
Roussev,I., Galsgaard,K., Erd\'{e}lyi,R. \& Doyle,J.G., 2001b, A\&A, 375, 228

\bibitem[Roussev 1]{R3}
Roussev,I., Doyle,J.G., Galsgaard,K. \& Erd\'{e}lyi, R., 2001c, A\&A, 380, 719

\bibitem[Sato \& Hayashi 1979]{S-H}
Sato,T. \& Hayashi,T. 1979, Phys.Fluids, 22, 1189

\bibitem[Semenov 1992]{SKLRHB}
Semenov,V.S., Kubyshkin,I.V., Lebedeva,V.V., Rijnbeek,R.P., Heyn,M. \& Biernat,H.K., 1992, Planet. Space Sci., 40, 63

\bibitem[Sonnerup 1970]{Son}
Sonnerup, B. U. \"{O}, J. Plasma Phys. 1970, 4, 161

\bibitem[Sweet 1958]{Swe}
Sweet, P. A. 1958, The neutral point theory of solar flares, in Electro-magnetic Phenomena in Cosmical Physics, ed. B.Lehnert (Cambridge University Press, London), 123

\bibitem[Tajima \& Shibata 1997]{T-S}
Tajima, T. \& Shibata, K. 1997, Plasma Astrophysics (Reading, Massachusetts: Addison-Wesley), 223

\bibitem[Tsuda \& Ugai 1977]{T-U}
Tsuda,T. \& Ugai,M. 1977, J. Plasma Phys., 18, 451

\bibitem[Ugai \& Tsuda 1977]{U-T1}
Ugai,M. \& Tsuda,T. 1977, J. Plasma Phys., 17, 337

\bibitem[Ugai \& Tsuda 1979]{U-T2}
Ugai,M. \& Tsuda,T. 1979, J. Plasma Phys., 22, 1

\bibitem[Ugai 1999]{Uga2}
Ugai,M. 1999, Phys. Plasmas, 6, 1522

\bibitem[Vasyliunas 1975]{Vas}
Vasyliunas,V.M. 1975, Rev.Geophys., 13, 303

\bibitem[Yokoyama 2001]{Yok}
Yokoyama,T., Akita,K., Morimoto,T., Inoue,K., Newmark, J., 2001, ApJ, 546L, 69

\end{thebibliography}
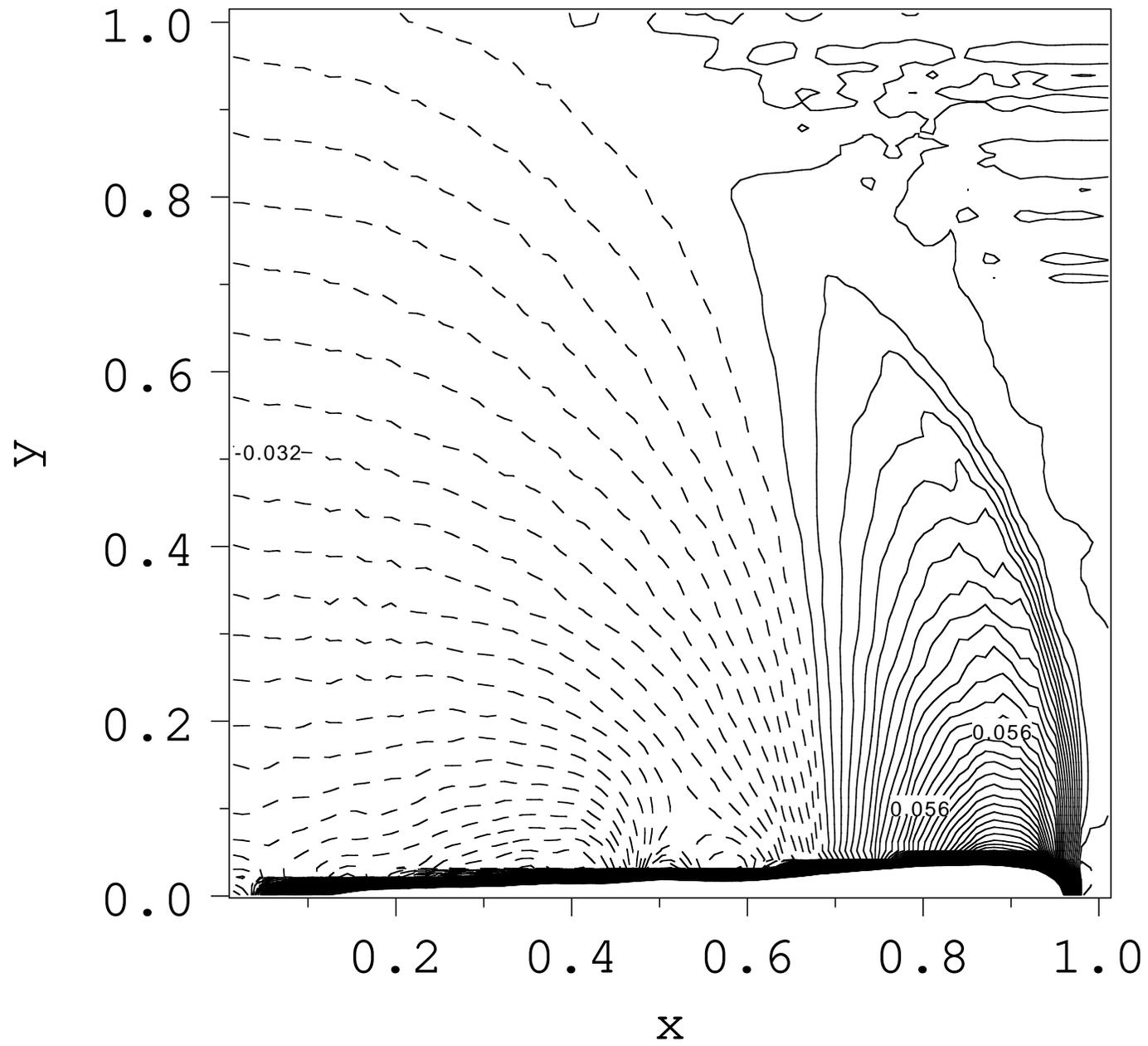
\end{document}